\documentclass[a4paper,sort&compress]{jpconf}

\usepackage{graphicx}
\usepackage[usenames,dvipsnames]{color}
\usepackage{bm}
\usepackage{amsmath}
\usepackage[utf8]{inputenc}
\usepackage{cite}

\usepackage{hyperref}
\hypersetup{
  colorlinks=true,
  linkcolor=blue,
  citecolor=blue,
  urlcolor=blue
}

\definecolor{darkblue}{RGB}{0,0,196}

\begin{document}
\title{Phenomenological predictions of 3+1d anisotropic hydrodynamics}

\author{Mohammad Nopoush and Michael Strickland}

\address{Department of Physics, Kent State University, Kent, OH 44242 United States}

\author{Radoslaw Ryblewski}

\address{
Institute of Nuclear Physics, Polish Academy of Sciences, PL-31342 Krakow, Poland}

\ead{mnopoush@kent.edu}

\begin{abstract}
We make phenomenological predictions for particle spectra and elliptic flow in heavy-ion collisions using 3+1d anisotropic hydrodynamics (aHydro) including the effects of both shear and bulk viscosities. The dynamical equations necessary are derived by taking moments of the Boltzmann equation allowing for three distinct (diagonal) momentum-space anisotropy parameters. The formulation is based on relaxation-time approximation for the collisional kernel and a lattice-QCD-based equation of state. Evolving the system to late times, we calculate particle production using THERMINATOR 2, modified to account for an ellipsoidal distribution function. We obtain particle spectra for different particle species such as pions, kaons, and protons, and elliptic flow $v_2$ as a function of centrality, transverse momentum, and rapidity. In our model, we have four free parameters, i.e. freeze-out temperature, initial central energy density, initial momentum-space anisotropies, and shear viscosity to entropy density ratio. Using a multidimensional fit to LHC experimental data, we make a preliminary extraction of these parameters. We find reasonable agreement between 3+1d aHydro and available experimental data for $\eta/s\sim 0.23$.
\end{abstract}

\section{Introduction}

In recent years, hydrodynamics has enabled us to enhance our understanding of the space-time evolution of the quark-gluon plasma (QGP) created in ultra-relativistic heavy-ion collisions. Different schemes of hydrodynamics such as ideal hydrodynamics and later viscous hydrodynamics (vHydro) have been able to describe the evolution of QGP quite well~\cite{Jeon:2016uym}. However, the existence of large momentum-space anisotropies at early times and near the transverse edges of QGP is numerically challenging and can cause standard vHydro schemes to break down. Anisotropic hydrodynamics (aHydro) addresses this issue by taking into account the inherent momentum-space anisotropy of the QGP at leading order in a reorganized hydrodynamic expansion~\cite{Florkowski:2010cf,Martinez:2010sc,Bazow:2013ifa,Tinti:2013vba,Nopoush:2014pfa,Tinti:2015xwa,Bazow:2015cha,Strickland:2015utc,Molnar:2016vvu,Molnar:2016gwq}. Importantly, aHydro has been shown to agree better with exact solution of the Boltzmann equation than vHydro models \cite{Florkowski:2013lza,Florkowski:2013lya,Denicol:2014tha,Denicol:2014xca,Nopoush:2014qba,Heinz:2015gka}. In this paper, we study the phenomenology of the QGP using a 3+1d aHydro formalism. Evolving the system to late times, we calculate the resulting identified particle spectra and elliptic flow generated using a fixed energy density freeze-out prescription. Our preliminary results show a reasonable agreement with the LHC experimental data for 2.76 $\rm TeV$ Pb$+$Pb collisions.  

\section{Setup and background}

Leading order anisotropic hydrodynamics (aHydro) is based on an anisotropic distribution function of the form \cite{Nopoush:2014pfa,Martinez:2012tu}
\begin{equation}
f(x,p)\equiv f_{\rm iso}\Big(\frac{\sqrt{p^\mu \Xi_{\mu\nu}p^\nu}}{\lambda}\Big)\,,
\label{eq:dist}
\end{equation}
with $\lambda$ being a temperature-like scale and $\Xi_{\mu\nu}\equiv u_\mu u_\nu+\xi_{\mu\nu}-\Phi \Delta_{\mu\nu}$ being the anisotropy tensor, which parametrizes the anisotropic form of distribution function. In this relation, $u_\mu$ is fluid four-velocity, $\xi_{\mu\nu}$ is a symmetric traceless anisotropy tensor which obeys $u_\mu \xi^{\mu\nu}=0$, $\Phi$ is the bulk degree of freedom, and $\Delta_{\mu\nu}\equiv g_{\mu\nu}-u_\mu u_\nu$ is the transverse projection operator. Herein, the tensor $\xi_{\mu\nu}$ is taken to be diagonal, i.e. $\xi_{\mu\nu}={\rm diag}(0,{\boldsymbol \xi})$, where $\boldsymbol \xi$ are anisotropy parameters in different spatial directions. This form contains information about both shear and bulk corrections at leading order \cite{Nopoush:2014pfa}. In what follows, we assume the distribution to be of Boltzmann form, i.e. $f_{\rm iso}(p)\equiv \exp{(-p)}$.
\section{Dynamical Equations}
In order to derive the dynamical equations, one needs to set up the basis vectors in the lab frame (LF). To do so, one needs to perform a set of tensor  transformations including a boost along and a rotation around the longitudinal beam direction, followed by a transverse boost which connects the local rest frame to the LF \cite{Florkowski:2010cf,Martinez:2012tu}. The dynamical equations are obtained by taking moments of the Boltzmann equation
\begin{equation}
p^\mu \partial_\mu f=-{\cal C}[f]\,,
\end{equation}
 assuming the anisotropic form for the distribution function (\ref{eq:dist}) and then taking the relevant projections using the LF basis vectors. The zeroth and the first moments lead to evolution equations for the particle four-current $J^\mu$ and the energy-momentum tensor $T^{\mu\nu}$. Taking higher moments, one obtains the dynamical  equations for hydrodynamics quantities of higher rank, i.e. the second moment of the Boltzmann equations provides information about microscopic dissipation. The result is a closed set of dynamical equations,
\begin{equation}
\partial_\mu J^\mu=(n_{\rm eq}-n)/\tau_{\rm eq}\,,\quad
\partial_\mu T^{\mu\nu}=0\,,\quad
\partial_\mu {\cal I}^{\mu\nu\lambda}=u_\mu({\cal I}^{\mu\nu\lambda}_{\rm eq}-{\cal I}^{\mu\nu\lambda})/\tau_{\rm eq}\,,
\end{equation}
which is obtained based on the relaxation-time approximation for the collisional kernel. In the equations above, $J^\mu$, $T^{\mu\nu}$, and ${\cal I}^{\mu\nu\lambda}$ are the first, second, and third moments of the distribution function, respectively. Note that the $n^{\rm th}$-moment of distribution function is defined by $N_{\rm dof}\int d^3 p/((2\pi)^3E) p^{\mu_1}\cdot\cdot\cdot p^{\mu_n}f(x,p)$, with $N_{\rm dof}$ being number of degrees of freedom. Herein, we obtain four equations from the first moment and three from the second moment. In addition, the conservation of energy provides us with an extra equation which allows us to compute the local effective temperature from the non-equilibrium energy density (Landau-matching).

\section{\protect{a}Hydro equation of state}
Implementing a realistic equation of state (EoS) in the framework of aHydro is conceptually challenging since the EoS is a relation between the energy density and pressure in isotropic equilibrium. We have devised two approaches to deal with this issue. In the quasiparticle approach, we assume that the system is comprised of quasiparticles with a temperature-dependent mass which is fit to the lattice QCD data \cite{Alqahtani:2015qja,Alqahtani:2016rth}. In the second approach, called the standard approach, we obtain the necessary dynamical equations in the conformal limit ($m\rightarrow 0$). In this limit, the components of $T^{\mu\nu}$ multiplicatively factorize and one can then connect the isotropic parts of energy density and pressures using a realistic EoS \cite{Nopoush:2015yga}. In this work, we have used the standard approach. The realistic equilibrium EoS used herein is taken from Krakow parametrization of lattice data \cite{Chojnacki:2007jc}.

\section{\protect{a}Hydro freezeout}
Evolving to late times, the system undergoes a crossover from quarks and gluons to hadronic degree of freedom and later on the kinetic freezeout. In order to compare the result of our hydrodynamics model to experimental data, one needs to calculate the differential particle spectra at freezeout. This procedure is done by constructing a constant energy density  hypersurface, defined through an effective temperature $T_{\rm FO}={\cal E}^{-1}({\cal E}_{\rm FO})$. Then, by computing for the particles which cross this hypersurface, one can determine the number of hadrons produced in heavy-ion collisions at the freezeout. For this purpose, we apply a generalized Cooper-Frye formula, modified for aHydro \cite{Alqahtani:2016rth}. 

\section{Results}
In this section, we present our preliminary results and compare them to LHC experimental data for 2.76 $\rm TeV$ Pb$+$Pb collisions \cite{Abelev:2013vea,Abelev:2014pua}. The results are generated for many hadronic event samples at freezeout. In the plots shown herein we sampled 20,100 hadronic events using THERMINATOR 2 \cite{Chojnacki:2011hb}. The system is initialized using an smooth optical Glauber profile and assumed to be initially isotropic in momentum space. In our model, we have some free parameters including initial central temperature $T_0$, shear viscosity to entropy density ratio $\eta/s$, initial momentum-space anisotropies of the system, and the freezeout temperature $T_{\rm FO}$. By fitting to the experimental data and determining the best fit, we have estimated the appropriate values for these parameters. Based on our model, the best fit parameters values are: $T_0=$ 0.56 GeV, $T_{\rm FO}=$ 0.13 GeV, and $\eta/s=3/4\pi$. 

\begin{figure}[t]
\centerline{
\includegraphics[width=0.48\linewidth]{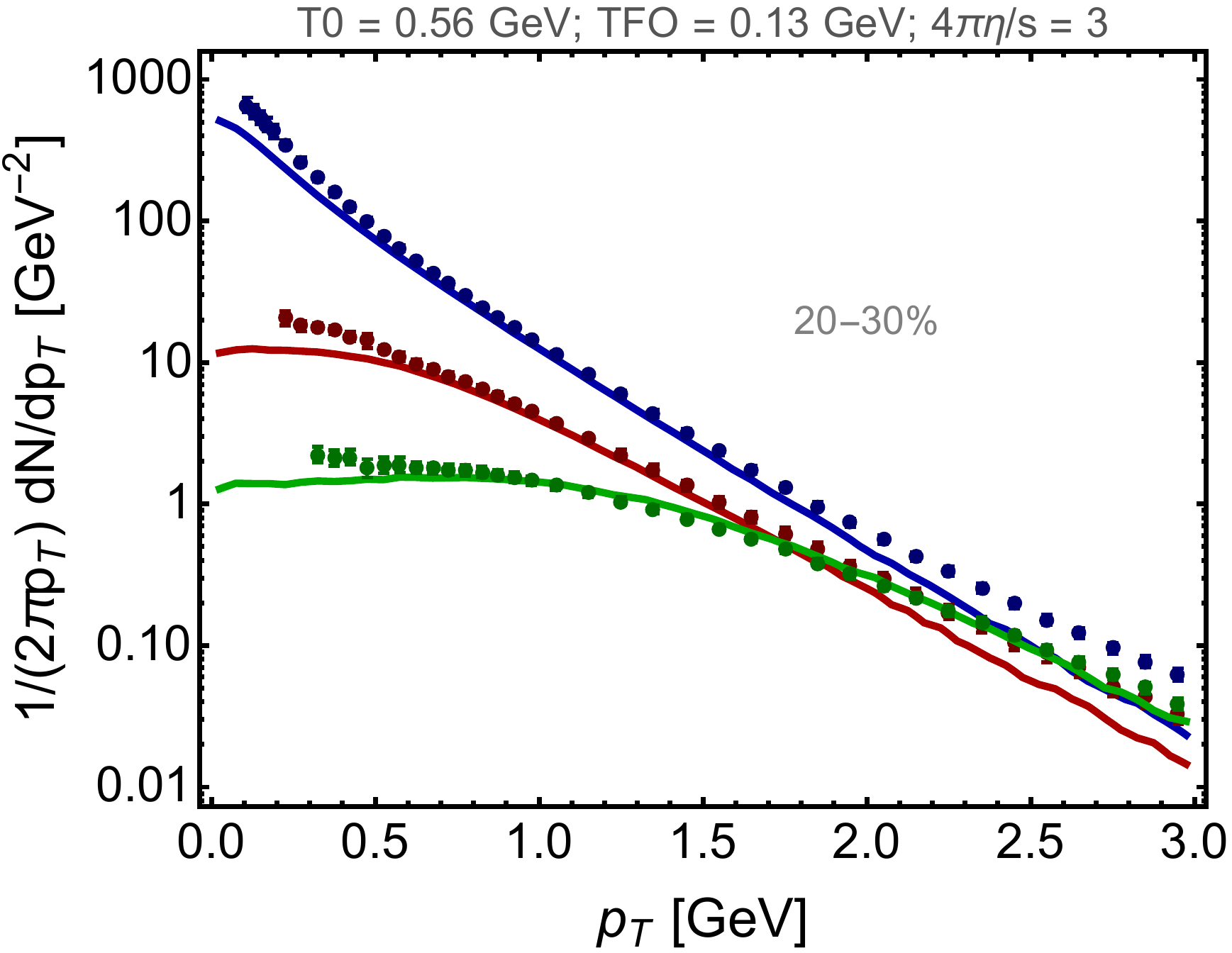}
\includegraphics[width=0.47\linewidth]{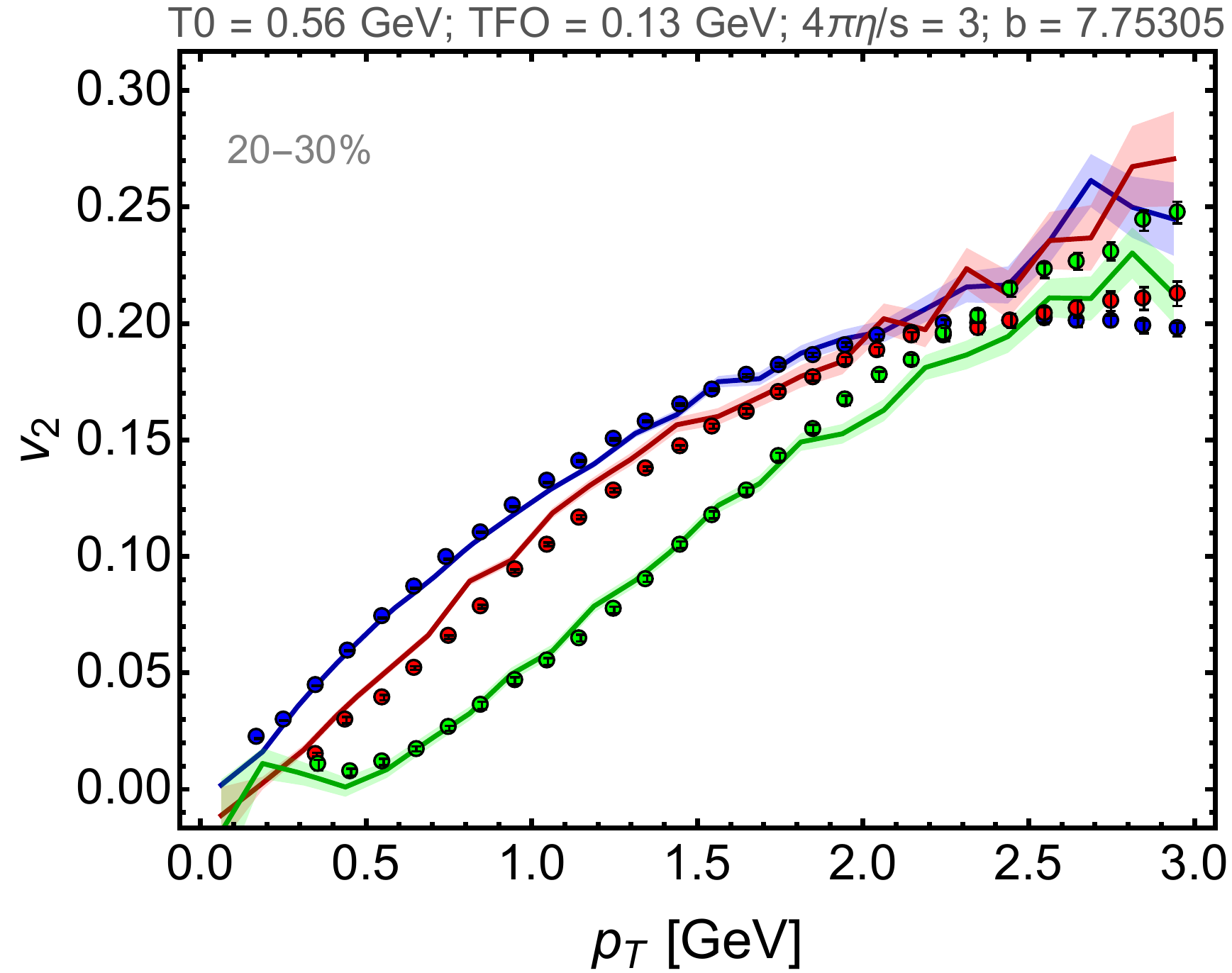}
}
\caption{In this figure, we compare aHydro predictions to experimental data for $\pi^+$ (blue), $K^+$ (red), and $p$ (green). The left panel shows the particle spectra and the right panel shows $v_2$, both as a function of transverse momentum $p_T$ for the 20-30$\%$ centrality class. The best fit corresponded to $T_0=$ 0.56 GeV, $T_{\rm FO}=$ 0.13 GeV, and $4\pi\eta/s=3$. Note that the experimental error bars shown are statistical only.}
\label{fig1}
\end{figure}

In Fig.~\ref{fig1}, we have plotted our results for LHC 2.76 $\rm TeV$ Pb$+$Pb collisions for 20-30$\%$ centrality class. In our plots, the results for charged pions (blue), kaons (red) and protons (green) are presented. The shaded bands are the statistical error associated with the THERMINATOR 2 Monte-Carlo sampling of the hypersurface. The left panel shows the $p_T$-differential particle spectra and the right panel shows $v_2$, both as a function of transverse momenta $p_T$. For $p_T<2\:\rm GeV$ the $v_2$ fit quality seems very good. Our results follow the experimental data not only in magnitude but also in curvature. For $p_T>2\:\rm GeV$ the fluctuations are significant because of the fewer number of events in this range of momentum. While the fit to $v_2$ seems quite promising, for very small $p_T$ our model underestimates the spectra, as can be seen from the left panel of Fig.~\ref{fig1}. We are currently investigating the cause of this discrepancy.

\section{Conclusions and outlooks}
In conclusion, we have implemented a 3+1d aHydro code for studying the quark-gluon plasma created in ultra-relativistic heavy-ion collisions. This model can be used to study particle spectra, flow coefficients, and other quark-gluon plasma experimental observables. By fitting to experimental data, we have managed to extract a preliminary estimation for some unknown parameters including initial central temperature, freezeout temperature, and shear viscosity to entropy density ratio. Our estimation for $\eta/s\simeq 0.23$ is quite different from a recent estimation by viscous hydrodynamics, $\eta/s\simeq 0.095$ \cite{Ryu:2015vwa}. In order to improve the accuracy of the model for the future studies, one needs to consider fluctuating initial conditions for the aHydro evolution.  Fluctuating initial conditions are already implemented in the 3+1d aHydro code, however, for this preliminary study, we restricted our attention to a smooth optical Glauber profile.  Looking forward, one also can take into account off-diagonal components of anisotropy tensor and the effect of chemical potential(s) in the anisotropic distribution function.

\section{Acknowledgements}
M.N. and M.S.  were supported by the U.S. Department of Energy, Office of Science, Office of Nuclear Physics under Award No. DE-SC0013470. R.R. was supported by Polish National Science Center Grant DEC-2012/07/D/ST2/02125.

\section*{References}

\bibliography{HQ2016-Nopoush}

\providecommand{\newblock}{}
\begin{thebibliography}{10}
\expandafter\ifx\csname url\endcsname\relax
  \def\url#1{{\tt #1}}\fi
\expandafter\ifx\csname urlprefix\endcsname\relax\def\urlprefix{URL }\fi
\providecommand{\eprint}[2][]{\url{#2}}

\bibitem{Jeon:2016uym}
Jeon S and Heinz U 2016 {Introduction to Hydrodynamics} {\em Quark-Gluon Plasma
  5\/} ed Wang X~N pp 131--187

\bibitem{Florkowski:2010cf}
Florkowski W and Ryblewski R 2011 {\em Phys. Rev.\/} {\bf C83} 034907
  (\textit{Preprint} \eprint{1007.0130})

\bibitem{Martinez:2010sc}
Martinez M and Strickland M 2010 {\em Nucl. Phys.\/} {\bf A848} 183--197
  (\textit{Preprint} \eprint{1007.0889})

\bibitem{Bazow:2013ifa}
Bazow D, Heinz U~W and Strickland M 2014 {\em Phys. Rev.\/} {\bf C90} 054910
  (\textit{Preprint} \eprint{1311.6720})

\bibitem{Tinti:2013vba}
Tinti L and Florkowski W 2014 {\em Phys. Rev.\/} {\bf C89} 034907
  (\textit{Preprint} \eprint{1312.6614})

\bibitem{Nopoush:2014pfa}
Nopoush M, Ryblewski R and Strickland M 2014 {\em Phys. Rev.\/} {\bf C90}
  014908 (\textit{Preprint} \eprint{1405.1355})

\bibitem{Tinti:2015xwa}
Tinti L 2016 {\em Phys. Rev.\/} {\bf C94} 044902 (\textit{Preprint}
  \eprint{1506.07164})

\bibitem{Bazow:2015cha}
Bazow D, Heinz U~W and Martinez M 2015 {\em Phys. Rev.\/} {\bf C91} 064903
  (\textit{Preprint} \eprint{1503.07443})

\bibitem{Strickland:2015utc}
Strickland M, Nopoush M and Ryblewski R 2016 {\em Nucl. Phys.\/} {\bf A956}
  268--271 (\textit{Preprint} \eprint{1512.07334})

\bibitem{Molnar:2016vvu}
Molnar E, Niemi H and Rischke D~H 2016 {\em Phys. Rev.\/} {\bf D93} 114025
  (\textit{Preprint} \eprint{1602.00573})

\bibitem{Molnar:2016gwq}
Molnar E, Niemi H and Rischke D~H 2016  (\textit{Preprint} \eprint{1606.09019})

\bibitem{Florkowski:2013lza}
Florkowski W, Ryblewski R and Strickland M 2013 {\em Nucl. Phys.\/} {\bf A916}
  249--259 (\textit{Preprint} \eprint{1304.0665})

\bibitem{Florkowski:2013lya}
Florkowski W, Ryblewski R and Strickland M 2013 {\em Phys. Rev.\/} {\bf C88}
  024903 (\textit{Preprint} \eprint{1305.7234})

\bibitem{Denicol:2014tha}
Denicol G~S, Heinz U~W, Martinez M, Noronha J and Strickland M 2014 {\em Phys.
  Rev.\/} {\bf D90} 125026 (\textit{Preprint} \eprint{1408.7048})

\bibitem{Denicol:2014xca}
Denicol G~S, Heinz U~W, Martinez M, Noronha J and Strickland M 2014 {\em Phys.
  Rev. Lett.\/} {\bf 113} 202301 (\textit{Preprint} \eprint{1408.5646})

\bibitem{Nopoush:2014qba}
Nopoush M, Ryblewski R and Strickland M 2015 {\em Phys. Rev.\/} {\bf D91}
  045007 (\textit{Preprint} \eprint{1410.6790})

\bibitem{Heinz:2015gka}
Heinz U, Bazow D, Denicol G~S, Martinez M, Nopoush M, Noronha J, Ryblewski R
  and Strickland M 2016  (\textit{Preprint} \eprint{1509.05818})

\bibitem{Martinez:2012tu}
Martinez M, Ryblewski R and Strickland M 2012 {\em Phys. Rev.\/} {\bf C85}
  064913 (\textit{Preprint} \eprint{1204.1473})

\bibitem{Alqahtani:2015qja}
Alqahtani M, Nopoush M and Strickland M 2015 {\em Phys. Rev.\/} {\bf C92}
  054910 (\textit{Preprint} \eprint{1509.02913})

\bibitem{Alqahtani:2016rth}
Alqahtani M, Nopoush M and Strickland M 2016  (\textit{Preprint}
  \eprint{1605.02101})

\bibitem{Nopoush:2015yga}
Nopoush M, Strickland M, Ryblewski R, Bazow D, Heinz U and Martinez M 2015 {\em
  Phys. Rev.\/} {\bf C92} 044912 (\textit{Preprint} \eprint{1506.05278})

\bibitem{Chojnacki:2007jc}
Chojnacki M and Florkowski W 2007 {\em Acta Phys. Polon.\/} {\bf B38}
  3249--3262 (\textit{Preprint} \eprint{nucl-th/0702030})

\bibitem{Abelev:2013vea}
Abelev B {\em et~al.\/} (ALICE) 2013 {\em Phys. Rev.\/} {\bf C88} 044910
  (\textit{Preprint} \eprint{1303.0737})

\bibitem{Abelev:2014pua}
Abelev B~B {\em et~al.\/} (ALICE) 2015 {\em JHEP\/} {\bf 06} 190
  (\textit{Preprint} \eprint{1405.4632})

\bibitem{Chojnacki:2011hb}
Chojnacki M, Kisiel A, Florkowski W and Broniowski W 2012 {\em Comput. Phys.
  Commun.\/} {\bf 183} 746--773 (\textit{Preprint} \eprint{1102.0273})

\bibitem{Ryu:2015vwa}
Ryu S, Paquet J~F, Shen C, Denicol G~S, Schenke B, Jeon S and Gale C 2015 {\em
  Phys. Rev. Lett.\/} {\bf 115} 132301 (\textit{Preprint} \eprint{1502.01675})

\end{thebibliography}

\end{document}